\documentclass{svproc}
\usepackage{url}

\usepackage{graphicx}
\newcommand{\GPN}{GPN\textsuperscript{\textregistered}}

\begin{document}
\mainmatter              
\title{Machine Learning Prediction of Gamer's Private Networks (\GPN s)}
\titlerunning{ML Analysis and Prediction of \GPN s}  
%
\author{Chris Mazur\inst{1} \and Jesse Ayers\inst{1}, Ga\'etan Hains\inst{2} \and Youry Khmelevsky\inst{1}}
\authorrunning{Chris Mazur, Jesse Ayers, Ga\'etan Hains, and Youry Khmelevsky} 
%
\tocauthor{}
\institute{Computer Science, Okanagan College, Kelowna, BC V1Y 4X8, Canada,\\
\email{ykhmelevsky@okanagan.bc.ca},\\ WWW home page:
\texttt{http://www.okanagan.bc.ca/cosc/}
 \and
LACL, Univ. Paris-Est Cr\'eteil, France
}

\maketitle              

\begin{abstract}
The Gamer's Private Network (\GPN) is a client/server technology created by WTFast for making the network performance of online games faster and more reliable. \GPN s use middle-mile servers and proprietary algorithms to better connect online video-game players to their game's servers across a wide-area network.

Online games are a massive entertainment market and network latency is a key aspect of a player's competitive edge. This market means many different approaches to network architecture are implemented by different competing companies and that those architectures are constantly evolving. Ensuring the optimal connection between a client of WTFast and the online game they wish to play is thus an incredibly difficult problem to automate.

Using machine learning, we analyzed historical network data from \GPN{} connections to explore the feasibility of network latency prediction which is a key part of optimization. Our next step will be to collect live data (including client/server load, packet and port information and specific game state information) from \GPN{} Minecraft servers and bots. We will use this information in a Reinforcement Learning model along with predictions about latency to alter the clients' and servers' configurations for optimal network performance.

These investigations and experiments will improve quality of service and reliability for \GPN{} systems.

\end{abstract}

\section{Introduction}
\setlength{\fboxrule}{2pt} 

The ``GPNPerf" (2014--2015, 2019--2021) project has been using various simulations, data analysis techniques, and Machine Learning (ML) models to investigate game connections over \GPN. 

The findings of this paper represents recent efforts (January--April 2020) in predicting the Round Trip Time (RTT) of a connection. The experiments include a Support Vector Machine (SVM) ML model to accurately predict a connection's round trip time (RTT) to within 30 milliseconds and the Neural Network (NN) ML model to make predictions on ping with a mean absolute error of 12.85 milliseconds.

These prediction models continue to be improved as the quality and quantity of data we work with increases. They provide the groundwork for the creation of a specific optimization example using Minecraft servers and bots which will be built with generalization to other games in mind. 

We wish to understand how network traffic evolves in order to better predict it. To do so, we look at the network statistics and measure them against some software and hardware variables \cite{Ward2017GamingND}.

In the following sections we will briefly describe specific game traffic connection types, then existing works and latency prediction with neural networks. In the following section we describe latency prediction with support vector machines (SVM). In the Future Work section we explain our future research goals using neural network with home developed bots for online game ``Minecraft". 

\section{Specific Game Traffic Connection Types}
\label{SpecificGametrafficConnectionTypes}

A core concern of the GPNPerf2 project is the analysis of \GPN{} traffic to define the necessary characteristics of a network traffic latency prediction model \cite{Ward2017GamingND}. Latency reduction is both the heart of WTFast's business and the key qualify feature of games networks. In their paper {\em QoE and Latency Issues in Networked Games} \cite{saldana2017qoe}, Saldana and Suznjevic confirm the necessity of low latency, even above that of bandwidth throughput, for player engagement in almost every kind of online game (QoE is Quality of Experience).

Measurement and modeling objectives for the GPNPerf2 project, based on concepts and results from \cite{saldana2017qoe}, will be discussed below.

\paragraph{Game genres i.e. types of online games} \cite{Ward2017GamingND}:
\begin{itemize}
	\item First Person shooters (FPS).
	\item Massively Multiplayer Online Role Playing Games (MMORPG) are games where thousands of players merge with artificial entities into a complex virtual world. 
	\item Real Time Strategy (RTS). 
	\item Multiplayer Online Battle Arena (MOBA) games. 
	\item Sports games that simulate car racing or team sports. 
\end{itemize} 
			
Authors in \cite{saldana2017qoe} quote existing surveys that for FPS games, a one-way delay of 80ms can be acceptable for most game users \cite{Ward2017GamingND}.
			
For MMORPG games, players started rating the game quality from ``excellent" to ``good" when one-way latency raised above 120ms \cite{Ward2017GamingND}. 

Geographical location of servers is correlated with latency for obvious reasons of transmission delays \cite{Ward2017GamingND}.

\section{Existing works}
    Consistent sub-second reaction time is a key consideration for intuitive PC frameworks \cite{Doherty1982}. For most video games this is an undeniable prerequisite that cutting-edge equipment has fulfilled, regardless of the increase in those games' demands for more complex interactions and improved graphical quality.
        
    A {\em video game network} is a distributed set of ``apparatus which [is] capable of exhibiting an interactive single identity game," as defined in a patent dated 1986 \cite{Sitrick1986}. The requirements for response time are even more stringent in this context and in addition to inevitable network latencies, ``the on-line service's computers themselves introduce latencies, typically increasing as the number of active users increases" \cite{Perlman1996}. The work completed by the previous iteration of the project \cite{Ward2017GamingND} consisted of a test examination of the conditions for fulfilling this key prerequisite, particularly in low and unsurprising reaction time for a game system aimed toward a varying quantity of players.
        
    The past fifteen years have seen a developing enthusiasm for handling this issue. A few analysts like Iimura \cite{Iimura2004}, Jardine \cite{Jardine2008} and co-creators have suggested shared structures for multiplayer online video games with the expectation of decreasing the data transfer capacity and preparing perquisites on servers. This has the potential of better scaling, but ``opens the game to additional cheating, since players are responsible for distributing events and storing state". Pellegrino et al. \cite{Pellegrino2003} have then proposed a hybrid architecture called P2P with central arbiter. The transmission capacity necessities on the router are lower than the server of a unified design. In the same way as other non-utilitarian properties of online systems (security, adaptability, unwavering quality and so on) the decision among centralization and appropriation isn't one that can be offered a complete response. Ward et al. \cite{Ward2017GamingND} focused on a legitimately unified architecture that has capability for consistency and versatility of the server and router execution.
        
    Other studies \cite{Ghosh2008} have contemplated similar execution issues within the sight of versatile player hubs. Regardless of its significance for the future, this line of focus shows it is even less developed than the P2P approach. Performance problems studied by \cite{Ghosh2008} in the presence of mobile player nodes reflect this. Despite its clear importance for the future, this mobile architectures appear even less mature than the P2P approach.
        
    The implementation of a Latency Management System (LMS) is a solution that gets implemented into game networks. Inconsistent latency poses a concern for developers who wish to create a fair game environment where all players can share in a similar experience. However, due to poor network traffic conditions, how players interact with the game and one another can be significantly hindered by the traffic conditions of others. To compensate for this issue, there are many different mechanisms that exist to conceal the problems created by varied network conditions which can be implemented depending on the needs of the developer \cite{saldana2017qoe}. These mechanisms take packet loss, jitter, and server delay into account to attempt to minimize the adverse effects that can occur. For example, ``shot behind cover" lag which can be encountered when a player with sufficiently poor network condition interacts with where another player was rather than where they currently are in relation to their own game client. In such a case, the game acknowledges the client-side interactions of the lagging player \cite{Lee:2018:EEM:3204949.3204971} which leads to poor experiences for players.
        
    One LMS solution is to group players with similar network conditions together \cite{ng2019systems}. This allows players with stable network conditions to receive an optimal experience as the affects of latency in highly competitive environments can heavily affect performance \cite{hoang2017lag}. Another solution is the prospect of optimizing data flow at the network level \cite{saldana2016effectiveness}. This solution can provide positive results for slower networks but risks stressing the unfairness between flows if network congestion is severe. These types of solutions go beyond what was hoped to address in their paper \cite{Ward2017GamingND}, but understanding the need for this information is important for the growth of gaming network solutions.
        
    ``A study of different first-person games shows that the client traffic is characterized by an almost constant packet and data rate" \cite{Faerber2004}. The study found that ``the average inter-packet time for client to server traffic to be 51ms for the game being studied". The bot system created in \cite{Ward2017GamingND} sends its action packets at 50ms intervals \cite{BotSyscon2015} in order to better observe how the networking design affects latency.
        
    As it was shown in \cite{Abdelkhalek:2003}, the ``bottleneck in the server is both game-related as well as network-related processing (about 50\%-50\%)". In \cite{Ward2017GamingND}, the examination done generally focused on the servers' exhibited improvement, the system traffic investigation \cite{visiogame2014}, and the execution of a custom bot for Minecraft \cite{BotSyscon2015}. During this exploration, the most noteworthy remaining task at hand for the CentOS 6.5 virtual server was examined by using a custom-created bot for Minecraft. 
        
    In a paper written by Jardine et al., \cite{Jardine2008} ``massively multiplayer online games with a client-server architectures and peer-to-peer game architectures" are investigated. The creators of these architectures built a hybrid game architecture to diminish game server data transfer capacity. In the paper written by Iimura et al. \cite{Iimura2004}, their findings included that creators even proposed to execute a zoned organization model for the multi-player online games attempting to lessen load on the game servers. A US 5956485 patent \cite{Perlman1996} portrays how to interface various remote players of online games on a conference phone line in a way which could lessen latency for the game players. This concept has been used and expanded upon with more modern internet technologies.
        
    Within the medium of online gaming, there is a strong continuing trend of Free-to-Play models, of which many different online games have millions of subscribers and hundreds of thousands of concurrent players \cite{gao2018energy,burger2016load,abdulazeez2018dynamic}. Due to the nature of how many of these games fragment their players for individual game sessions, a popular model discussed is the concept of a hybrid peer-to-peer (P2P) network architecture. This architecture ensures that players within reasonable proximity and network characteristics are paired together by a latency management system \cite{ng2019systems,garcia2018network} but also has the connection of the players rely on a centralized server to ensure game integrity. Within a fully P2P game environment, the absence of a centralized server results in all crucial data comes from the game host which has the potential risk of the host deliberately cheating or sending malicious data \cite{abdulazeez2016simulation} which can have varying degrees of consequence for all connected players. 
        
    Hybrid P2P servers rely on a large web of edge servers \cite{Plumb:2018:HNC:3235765.3235785} that take advantage of their large regional diversity to help minimize latency between the centralized servers and players, and reduce excess strain. 
        
    Cloud gaming, or gaming on demand, has become another popular alternative for creating network architectures for online games. MMOs are regularly turning to cloud gaming as a network solution as the number of concurrent players climbs into the hundreds of thousands \cite{gao2018energy}. Cloud gaming offers a scalable solution that handles large changes in players while helping manage the cost of bandwidth consumption \cite{slivar2019qoe}. This has become an increasingly popular medium for online games but comes with a certain level of stigma as the medium has been surveyed many times and customers are wary of the drawbacks \cite{khmelevsky2017stochastic}. The biggest risk involved when choosing to implement cloud gaming is that the Quality of Experience (QoE) comes with a large set of challenges stemming largely from challenges regarding latency \cite{mcmanus2019effects,chen2016inter}. Competitive online games that rely on minimizing latency and packet time often avoid cloud networks as even current Inter-player Delay Optimization (IDO) solutions only serve to reduce the perception of response delay from players rather than eliminating it \cite{cai2016survey,mcmanus2019effects}. However, cloud gaming strongly serves certain Massively Multiplayer Online (MMO) Games; while the experiences of players in First Person Perspective games suffer more from perceived latency issues, many other genres do not. 
        
%
    
\newpage

	\section{TraceRoutes Data and Markov Model}
   
In this section we describe our first model of game-network latency, the data it was applied to and the results that led us to review the modelling objectives, data and technique. 

This analysis had for ultimate objective the real-time prediction of latency spikes during the actual client-server routing. It was too ambitious but much was learned in the process. The data consisted of two files of traceroute measurements to measure and analyse the routing of messages between game client and game server. One file used normal internet (non-GPN in our jargon) and the other one used a GPN. Each measurement consisted of the following fields: 

\begin{itemize}
	\item IP Address, Latitude, and longitude, for the source and destination machine
	\item A list of up to 25 hop delays, measured in milliseconds
\end{itemize}

Each hop delay corresponds to a segment of route between client and server, their sum being the total latency. The GPN file contained 29584 measurements  between 108 source IP addresses and 214 destination IP addresses. The average number of hops per traceroute measurement was 13. The non-GPN file contained only 234 measurements (much larger files later confirmed the same statistics are described here) and involved 85 destination IP addresses from a unique one in our laboratory. The average number of hops per traceroute in it was 16. 

Our target for latency prediction was the occurrence of spikes i.e. relatively rare events in the sequences of hop delays. a hop-time that is above a certain threshold, initially 15ms. The absence of spikes is a measure of reliable and low-latency routing and the GPN clearly does guarantee this, but statistics make this observation clearer here is how. Most hop delays are in the low ms counts, or even sub-ms. In GPN traceroutes, spikes are very rare (they occur in less than 1\% of routes) but very large. Their distribution is between 61 and 160ms. For normal internet (non-GPN) the spikes are of shorter yet large delays, between 19 and 105ms, but much more frequent: they occur in 40\% of the traceroutes. 

Given the very negative impact of latency spikes, we hoped to predict them from a Markov model of the traceroute times series. To this end we defined a set of four “states” of a hop-delay time series as follows: 

\begin{center}
	\begin{tabular}{ | c | } 
		\hline
		[0=low] previous hop: 		[0, 0.5) ms \\
		\hline
		[1=avg]  previous hop : 	[0.5, 2.) ms  \\
		\hline
		[2=high]   previous hop :    	[2, 15) ms \\
		\hline
		[3=spike]   previous hop : 	[15, max) ms \\
	 	\hline
	\end{tabular}
\end{center}

Then we defined for each file a training subset of lines, the rest being the test set of lines. From the training sets we computed the following 4x4 Markov matrix of transitions probabilities.

\begin{center}
	\begin{tabular}{  c  } 
		M(i,j) = Pr[ State(t+1) = j |  State(t) = i ] 
	\end{tabular}
\end{center}

We applied this matrix to create a prediction of the instantaneous next states in the test time series, measuring the quality of prediction by the number of well-predicted spikes. We used between 10\% and 20\% of the traceroutes for training and tested our model on the 80-90\% others. The Markov matrix for GPN data is: 

\begin{center}
	\begin{tabular}{ |c|c|c|c| } 
		 \hline
		49.6\% & 50.1\% & 0.0\%  & 0.1\% \\	
		\hline 
		50.7\% &	49.2\% &	 0.0\% &	0.0\%L \\ 
		\hline
		33.3\% &	7.5\% & 	29.0\% &	30.1\%  \\ 
		\hline
		21.4\% &	 4.0\% &	20.8\% &	53.6\% \\ 
	 	\hline
	\end{tabular}
\end{center}

Here the first column contains the probabilities of having a low hop delay after each of the four possible previous hop delay “states”. The fourth column contains the probabilities of having a delay spike after each of the possible states etc. The Markov matrix for non-GPN data is: 

\begin{center}
	\begin{tabular}{ |c|c|c|c| } 
		 \hline
		49.6\% & 50.1\% & 0.0\%  & 0.1\% \\	
		 \hline
		70.2\% & 	9.7\%  & 	12.2\%  & 	7.7\% \\
		 \hline
		54.7\% & 	11.9\% & 	21.4\% & 	11.9\% \\
		 \hline
		52.1\% & 	26.0\% & 	18.4\% & 	3.2\% \\
		 \hline
		66.6\% & 	11.1\% & 	7.4\%	 & 14.8\% \\
	 	\hline
	\end{tabular}
\end{center}

If we multiply an initial uniform-probability vector (4 entries of 25\%) by successive powers of those matrices we obtain a general stable state distribution for each dataset. For the GPN data this limit distribution is: 

\begin{center}
	\begin{tabular}{ | c | c | c | c | }
		\hline 
		50.1\% &    49.5\% &     0.1\% &     0.2\%\\
		\hline
	\end{tabular}
\end{center}

and for the non-GPN data it is:

\begin{center}
	\begin{tabular}{ | c | c | c | c | }
		\hline 
		65.4\% &   12.4\% &    13.8\% &     8.2\% \\
		\hline
	\end{tabular}
\end{center}

This demonstrates that overall, about 8\% of all hop delays should be spikes in normal internet routing (vs 0.2\% with GPN). The 8\% value does not explain why spikes occurred in 40\% of our non-GPN routes but is non incoherent with it (we quantified this as a 4\% statistical error). The 0.2\% of GPN hops that are spikes is coherent (quantified as < 1\% statistical error) with the less than 1\% of GPN traceroutes that contain spikes. 

Then we tried to use our Markov model for a more demanding task: to predict one hop’s delay (state) from the previous one in its traceroutes time series. This was a failure due to the rarity of spike events, and possibly the fact that they are mostly unrelated to the previous hop times. Here are the detailed results of trying to predict the occurrence of a spike in the time series. For GPN data: 

\begin{center}
	\begin{tabular}{ | c | c | }
		\hline 
		0 false positive &	2 true positives \\
		\hline
		425 false negatives & 	305531 true negatives \\
		\hline
	\end{tabular}
\end{center}

which is an ``almost blind alarm system” i.e. one with no false alarms but many false negatives. For non-GPN data: 

\begin{center}
	\begin{tabular}{ | c | c | }
		\hline 
		0 false positive & 	0 true positives \\
		\hline
		236 false negatives & 	2870 true negatives \\
		\hline
	\end{tabular}
\end{center}

Here, the model is still perfect at predicting the 2870 non-spike hop times but misses all of the 236 spikes and predicting them as negatives.

In summary, the Markov model confirms and explains the superiority of GPN routing while quantifying the very high delays of its rare spike events. It fails to predict spike delays in real time but gives a good explanation of the distribution of latencies, including as a function of their position in the hop time series.
 
As a general conclusion we then modified our goals in two ways. A more modest analysis of the latency data: only round-trip time data and models, no hop details, but increased information in the model by using geolocation / date-time.

	\section{Latency Prediction with Neural Networks}

    \subsection{Overview}
    To use a neural network to attempt to predict the ping of the \GPN, a sample of data was used with TensorFlow as the machine learning library and the Keras API. This sample of data consisted of collected between March 2 2020 and March 10 2020. The data collected was 15158 records comparing the ping between \GPN{} and regular internet routing for players playing Final Fantasy XIV. 
    
    \subsection{Input Data}
    The data had its input fields reduced to the client's timestamp as well as the latitude and longitude values of the source, destination, and the one or two proxy servers that are part of the \GPN. Additionally, in order to capture the cyclical nature of time of day, a cosine transform was performed on the data set in order to convert it \cite{london:2016}.
        
    \begin{verbatim}
seconds_in_day = 24* 60 * 60
dataset['ClientTimestamp'] = numpy.cos(2*numpy.pi*
    dataset.ClientTimestamp/seconds_in_day)
    \end{verbatim}
        
\begin{figure}
        \centering
        \includegraphics[width=0.75\textwidth]{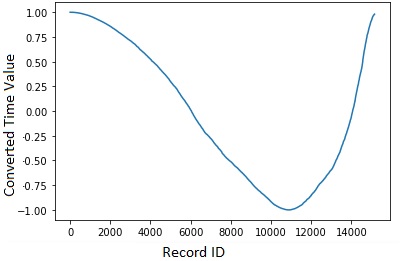}
        \caption{The time values after a cosine transform.}
        \label{fig:cosinetime}
    \end{figure}

    After this conversion, the result was graphed and can be seen in figure \ref{fig:cosinetime}. The reason this was done was to better convey to the neural network the time of day; both one minute before and after midnight are roughly the same part of the day, but numerically this is not captured by simply using the number of seconds elapsed that a usually parsed timestamp would return. The time of day relates to different amounts of internet traffic so it was important to include for prediction.
        
    \subsection{Model}
    The model was built as a Keras sequential model. The first layer was the input layer and thus it was shaped using the keys of the data set. Next there were two hidden layers of 128 neurons and finally an output layer with a single output. Both the input layer and the hidden layers operated under the ReLU activation function and the model used mean squared error for the loss function. Neither of these functions are unusual for a regression task.
        
    \subsection{Training and Results}
\begin{figure}
        \centering
        \includegraphics[width=0.75\textwidth]{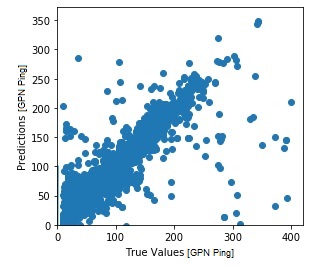}
        \caption{The comparison between true and predicted ping values for using the \GPN{} with Final Fantasy XIV.}
        \label{fig:predictionsnn}
    \end{figure}

    The model was then trained over 1000 epochs to predict the ping of the \GPN. 70\% of the data was used for training purposes. After training, the mean absolute error was reduced to 12.85 milliseconds. The results were graphed using the matplotlib library and can be seen in figure \ref{fig:predictionsnn}. The majority of data clumped closely to a linear line between the predicted and true values axes which suggested fairly accurate predictions.
    
    \subsection{Future for the Neural Network}
    As more data is collected, investigation into other data that could be used in the neural network will be conducted. For example, one hot encoding of categorical data such as country, continent, ISP, etc. Hopes are that this will assist in improving predictions; for example, two coordinates may be close to one another but are located in different countries with vastly different infrastructure which would affect the quality of a route.
        
    \newpage
        
    \section{Latency Prediction with Support Vector Machines}
	\label{SVM}

    	\subsection{Overview}
    		\noindent For this experiment, a Support Vector Machine (SVM) was used to predict total \GPN{} latency against a data set. An SVM is a supervised ML model that classifies data points in vector space using a hyperplane. The hyperplane which describes the separatation of classes of data with the most distance between the plane and the classes is then used to predict against the data, or to classify new data points. 
    
\begin{figure}[!h]
    		  \centering
    		  \includegraphics[width=0.75\textwidth]{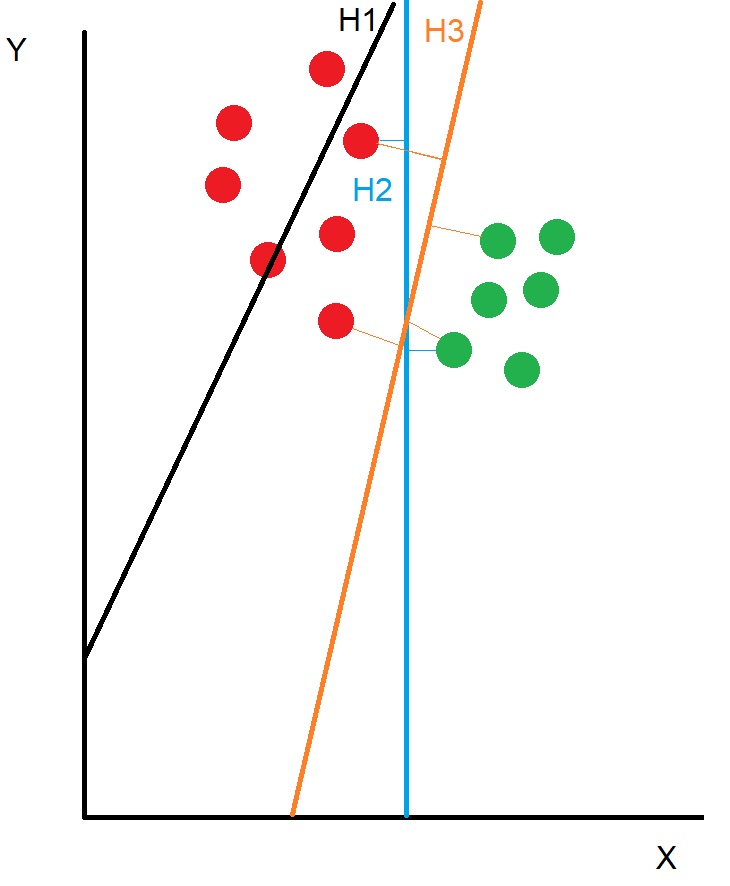}
    		  \caption{Example of an SVM separating two labelled sets}
    		  \label{fig:svm2set}
    		 \end{figure}

    		\noindent Figure \ref{fig:svm2set} shows the difference between a hyperplane that doesn't split the data (H1), a hyperplane that splits the data but that is not optimized with support vectors (H2), and a hyperplane that maximally separates the data with support vectors (H3).
    		
    		\noindent The classes to label the data in this experiment were derived from the \GPN{}RTT total latency. They are: Fastest (0-30ms), Fast (31-60ms), Normal-Fast (61-90ms), Normal-Slow (91-120ms), Slow (121-150ms), Slowest (151-180ms), and Unusable (181ms+). These categories were chosen based on videogame reporting statistics \cite{Claypool2006} which indicate pings under 50ms are most desirable, and that pings above 180ms see players begin to quit many types of games.
    
    		\subsection{Set Up}
    			\noindent The experiment was run using the R language version 3.6.3,  in the RStudio version 1.2.5033 environment. The specific SVM model used for this experiment was from R package e1071. The tuning function run on this model to determine the best parameters was also from the e1071 package. The caret package was used to generate confusion matrices and other reporting statistics from the SVM models. The geosphere package was used to transform the geolocation data into distance data.
    
    		\subsection{Input Data}
    			\noindent The data was gathered using a proprietary tool from WTFast. The data was gathered from two primary locations in Okanagan College and Burnaby, from December 20th 2019 to January 20th 2020. The data was mainly gathered during the morning and afternoon hours. 
    			
    			\noindent The data was initially in the form of .CSV files that were loaded into data frames. R natively supports data frames. There were two CSV files used; one which contained the IP addresses, latitude, and longitude of the Source and Destination, the names of the Proxy servers used, the timestamp of connection, and finally the \GPN{}RTT and Non\GPN{}RTT. The other .CSV file contained the names of the Proxy servers, along with their IP address, latitude, and longitude. 
    
    			\noindent These two .CSV files were used to create a third, merged .CSV for repeat tests, along with the data columns derived from the geosphere package: Distance (in kilometers) from Source to Destination, from Source to Proxy 1, from Proxy 1 to Proxy 2, and from Proxy 2 to Destination. 
    
    		\subsubsection{Data Columns Used}
    			\noindent Using a naive approach, all the data columns were kept and used as determinants for the label, except for the \GPN{} RTT, which was replaced by the label. 
    
    		\subsubsection{Data Columns Conversion}
    			\noindent R attempts to load columns into data frames from .CSV files as an appropriate data type. In this case, values were loaded as numeric or factor data types, both of which are usable by the SVM. Rows with na (null) values were removed using R's na.omit function, as SVMs do not interact well with na values. The Non-\GPN{}RTT, \GPN{}RTT columns, the various latitude, longitude, and distance, columns, and the IP and ID columns were left alone. The timestamp column was transformed to a Time-Series data type. 
    
    			\noindent The \GPN{}RTT column was used to derive the SpeedLabels column, based on the above mentioned categories, and a new data frame was created, replacing the \GPN{}RTT column with this column. From this data frame, five sets of two new data frames were created for training and testing the model, each as a random 80/20 split. These will be referred to as training sets 1-5 (containing random selections of 80\% of the data) and testing sets 1-5 (containing random selections of 20\% of the data). 
    
    		\subsection{Model}
    			\noindent The model being used is the SVM from the e1071 package. The parameters for this model relevant to this experiment are: 
    
    			\begin{itemize}
    			\item kernel: the formula describing the hyperplane. It can be linear (u\'*v),  polynomial ((gamma*u\'*v + coef0)\^degree), radial basis (exp(-gamma*|u-v|\^2)), or sigmoid (tanh(gamma*u'*v + coef0))
    			\item degree: the parameter needed for kernel of type polynomial (default: 3)
    			\item gamma: the parameter needed for all kernels except linear (default: 1/(data dimension))
    			\item coef0: the parameter needed for kernels of type polynomial and sigmoid (default: 0)
    			\item cost: affects how much the function changes when it encounters constraints violation; higher cost allows less change (default: 1)
    			\item cachesize: the maximum size of the SVM in MB (default: 40)
    			\item scale: whether or not the data is scaled internally, and how (default:  data is scaled to zero mean and unit variance)
    			\end{itemize}
    
    			\noindent The parameters for the model were determined by testing against the data in order to find the optimal model parameters. Testing for this experiment was performed using the ``tune" function, also from the e1071 package. The tune function is a grid search that compares the results of every combination of selected parameters in order to determine the best fit of the model for the data being used. It uses a 10-fold split on the training set to train and test the model, and further validates against a separate testing set. 
    
    			\noindent The SVM is being trained on a function derived from setting the SpeedLabel as the dependent variable, and every other column as the independent variables. The parameters being tuned are represented to the tune function as lists of available parameter options. 
    
    			\begin{itemize}
    			\item The values for the SVM's kernel were ``linear", ``polynomial", ``radial basis", and ``sigmoid". 
    			\item The values for the SVM's cost were a list of powers of 2, from 1/32 to 1024.
    			\item The values for the SVM's gamma were a list of powers of 2, from 1/32 to 4.
    			\item The values for the SVM's degree were a list of integers from 1 to 5 inclusive.
    			\item The values for the SVM's coef0 were 0.1, 0.5, 1, 2, 3 and 4.
    			\end{itemize}
    			Each SVM was given 200MB of cache size, and was told to scale the data frames internally between -1 and 1 to normalize the values with scale = TRUE. 
    
    			\noindent This tuning process is very compute intensive. You may wish to avoid it by using our reported values as follows:
    			\noindent  kernel = ``radial",  gamma = 1/32,  cost = 256,  scale = TRUE,  cachesize = 200
    
    		\subsection{Training}
    			\noindent With the parameters optimized, new svm models 1-5 were created and trained using the parameters. For model 1, it was trained on training set 1 and tested against testing set 1, and so forth. The test performed was using the predict function from the e1071 package, calling the model and test set as parameters. This generates a data frame of the model's predictions. Training an SVM model in this way against this data takes between 20 and 40 seconds on a modern i7 CPU, and the model converges before max iterations are reached. 
    
    			\noindent Although the e1071 library makes it difficult to access the number of iterations an SVM model has gone through and to alter the max number, the SVM it's based on, libSVM, uses ten million iterations as a base. Models trained during the tuning step of this experiment would provide an ``approaching max iterations" warning after roughly 2 minutes, which would indicate that our tuned models use between two and three hundred thousand iterations.
    
    		\subsection{Results}
    			\noindent The SVM model predictions are displayed using the caret package, which provides a function to produce confusion matrices. The confusion matrix is generated by providing the test set's actual labels and comparing them to the model's predictions. The data this confusion matrix provides is as follows:
    
    			\begin{itemize}
    			\item Accuracy : The percentage of label = prediction
    	                 	\item 95\% CI : The confidence interval range for our accuracy
    		    	\item No Information Rate : What we would expect the accuracy to be if we guessed based only on the distribution of the confusion matrix
       			\item P-Value [Acc $>$ NIR] : The likelihood we would get this result by chance
                                    \item Kappa : How reliable our accuracy measure is
    			\item Mcnemar's Test P-Value : This test is not available for our data
    			\end{itemize}
    
    			\noindent There is also a confusion matrix table that directly displays prediction against reference values. Values in the upper right are where the model predicts slower than the reference, and values in the lower left are where the model predicts faster than the reference. A representative example of this table and its associated output is provided below, and in table \ref{tab:svmresults1}.
    
    			\noindent \textbf{Overall Statistics} (Model 5)
    		           \begin{itemize}
    			\item Accuracy : 0.9278          
    		           \item 95\% CI : (0.9115, 0.9419)
    			\item No Information Rate : 0.2693          
    			\item P-Value [Acc $>$ NIR] : $<$ 2.2e-16       
    		           \item Kappa : 0.9123          
    			\item  Mcnemar's Test P-Value : NA 
    			\end{itemize}
    
    			\begin{table}[h]
    			\caption{Confusion Matrix and Statistics (Model 5)}
    			\label{tab:svmresults1} 
    			\centering
    				\label{Reference}
    				\begin{tabular}{ | p{2cm} | p{1cm} | p{.9cm} | p{1.2cm} | p{2cm} | p{1cm} | p{1.5cm} | p{1.8cm} | }  \hline
    					&	&	&	& Reference &	&	& 
    				\\ \hline
    				Prediction & 1 Fastest & 2 Fast & 3 N-Fast & 4 N-Slow & 5 Slow & 6 Slowest & 7 Unusable 
    				\\ \hline
    				1 Fastest & 33 & 4 & 0 & 1 & 0 & 0 & 0
    				\\ \hline
    				2 Fast & 4 & 135 & 12 & 0 & 1 & 0 & 0
    				\\ \hline
    				3 N-Fast & 0 & 10 & 174 & 3 & 0 & 0 &  0
    				\\ \hline
    				4 N-Slow & 0 & 0 & 4 & 77 & 11 & 0 & 0
    				\\ \hline
    				5 Slow & 0 & 0 & 0 & 1 & 290 & 4 & 2
    				\\ \hline 
    				6 Slowest &  0 & 0 & 0 & 0 & 13 & 17 & 6
    				\\ \hline
    				7 Unusable &   0 & 0 & 1 & 1 & 2 & 5 & 208
    				\\ \hline
    				 \end{tabular}
    			\end{table}
    
    		\noindent These results indicate that the model is most confused about the Slowest and Fastest categories, and is having a relatively large amount of trouble predicting against them. These results also indicate that the data being used to predict the label is extremely useful in doing so, when compared to the ``No Information Rate". 

\section{Future Work}
    Now that we have experiments that demonstrate that round trip time and ping can be predicted fairly accurately, we hope to expand this research into developing a solution for configuring the proxy servers needed in the \GPN. 
    
    A model such as the neural network that was trained could be used to accomplish this by using player and game server data to inquire on superior proxy sever configurations. However, we hope to take this another step further. Experiments to be performed in the summer of 2020 include setting up servers for the game ``Minecraft" where player bots will generate network traffic over the \GPN. 
    
    Our new neural network will then observe the server hardware and software data, the predicted \GPN{} RTT from our predictive models, and the \GPN{} configuration files made available to us from WTFast for their \GPN servers, in order to create a reinforcement learning model that will optimize each connection's RTT.
    
    If results from this experiment are successful, the next step will be to generalize the process to any game and server. 

\section{Conclusion}
    In these experiments, we have explored the ability to predict latency values when using a network that can have its routing configured like the \GPN. The results indicate that prediction is both possible and fairly accurate based on factors such as timestamp and coordinates.
    
     Our ability to predict the RTT provides a strong positive indicator that we will also be able to manipulate the \GPN s server-specific configuration in order to optimize performance in an intelligent and automated fashion between any number of clients and servers.

\section*{Acknowledgment}
\label{Acknowledgment}
    The research project results described in this paper were achieved with support from Computer Science department at Okanagan College, WTFast\textsuperscript{\textregistered}, and by the NSERC of Canada in 2014 (ARD1 465659 --- 14): ``GPN-Perf: Investigating performance of game private networks" and in GPN-Perf2 (2015--2017, 2019--2021).


%
%


\begin{thebibliography}{6}
%


\bibitem{BotSyscon2015}
Alstad, T.,  Dunkin, J.R., Detlor, S., French, B., Caswell, H., Ouimet, Z., Khmelevsky, Y.: Game Network Traffic Emulation by a Custom Bot. In: 2015 IEEE International Systems Conference (SysCon 2015) Proceedings (2015).

\bibitem{Doherty1982}
Doherty, W.J., Thadhani, A.J.: The economic value of rapid response time (IBM Technical Report GE20-0752-0) \url{http://www.vm.ibm.com/devpages/jelliott/evrrt.html}

\bibitem{Sitrick1986}
Sitrick, D.H.: Video Game Network. {U}nited {S}tates {P}atent Number 4,572,509.

\bibitem{Perlman1996}
Perlman, S.G.: Network Architecture to Support Multiple site real-time video games. {U}nited {S}tates {P}atent Number 5,586,257.

\bibitem{Ward2017GamingND}
Ward, B., Khmelevsky, Y., Hains, G., Bartlett, R., Needham, A., Sutherland, T.: Gaming network delays investigation and collection of very large-scale data sets. In: 2017 Annual IEEE International Systems Conference (SysCon) (2017).

\bibitem{Iimura2004}
Iimura, T.,Hazeyama, H., Kadobayashi, Y.: Zoned Federation of Game Servers: A Peer-to-peer Approach to Scalable Multi-player Online Games. In: Proceedings of 3rd ACM SIGCOMM Workshop on Network and System Support for Games (2004).

\bibitem{Jardine2008}
Jardine, J., Zappala, D.: A Hybrid Architecture for Massively Multiplayer Online Games. In: Proceedings of the 7th ACM SIGCOMM Workshop on Network and System Support for Games (2008).

\bibitem{Ghosh2008}
Ghosh, P., Basu, K., Das, S.K.: Improving end-to-end quality-of-service in online multi-player wireless gaming networks. In: Computer Communications (2008), vol. 31, pp. 2685--2698.

\bibitem{saldana2017qoe}
Saldana, J., Suznjevic, M.: QoE and Latency Issues in Networked Games. In: Handbook of Digital Games and Entertainment Technologies (2017), pp. 509--544.

\bibitem{Lee:2018:EEM:3204949.3204971}
Lee, S.W.K., Chang, R.K.C.: Enhancing the Experience of Multiplayer Shooter Games via Advanced Lag Compensation. In: Proceedings of the 9th ACM Multimedia Systems Conference (2018), pp. 284--293. \url{http://doi.acm.org/10.1145/3204949.3204971}

\bibitem{hoang2017lag}
Hoang, D.C., Doan, K.D., Hoang, L.T.: Lag of Legends: The Effects of Latency on League of Legends Champion Abilities. Worcester Polytechnic Institute (2017).

\bibitem{saldana2016effectiveness}
Saldana, J.: On the effectiveness of an optimization method for the traffic of TCP-based multiplayer online games. In: Multimedia Tools and Applications (2016), vol. 75 num. 24, pp. 17333--17374.

\bibitem{Abdelkhalek:2003}
Abdelkhalek, A., Bilas, A., Moshovos, A.: Behavior and Performance of Interactive Multi-Player Game Servers. In: Cluster Computing (2016), vol. 6 num. 4, pp. 355--366. \url{http://dx.doi.org/10.1023/A:1025718026938}

\bibitem{visiogame2014}
Alstad, T., Dunkin, J.R., Bartlett, R., Needham, A., Hains, G., Khmelevsky, Y.: Minecraft computer game simulation and network performance analysis. In: Second International Conferences on Computer Graphics, Visualization, Computer Vision, and Game Technology {(VisioGame 2014)} (2014).

\bibitem{gao2018energy}
Gao, Y., Wang, L., Xie, Z., Guo, W., Zhou, J.: Energy-Efficient and Quality of Experience-Aware Resource Provisioning for Massively Multiplayer Online Games in the Cloud. In: International Conference on Service-Oriented Computing (2018), pp. 854--869.

\bibitem{abdulazeez2018dynamic}
Abdulazeez, S.A., El Rhalibi, A.: Dynamic Load Balancing for Massively Multiplayer Online Games Using OPNET. In: International Conference on E-Learning and Games (2018), pp. 177--191.

\bibitem{ng2019systems}
Ng, F., Hannigan, J., Moon, D.: Systems and Methods for Managing Latency in Networked Competitive Multiplayer Gaming. (2019) Google Patents. US Patent App. 16/170,599.

\bibitem{garcia2018network}
Garcia, M.M., Woundy, R.M.: Network latency optimization. (2018) Google Patents. US Patent 9,998,383.

\bibitem{abdulazeez2016simulation}
Abdulazeez, S.A., El Rhalibi, A., Al-Jumeily, D.: Simulation of Massively Multiplayer Online Games communication using OPNET custom application. In: 2016 IEEE Symposium on Computers and Communication (ISCC), pp. 97--102.

\bibitem{Plumb:2018:HNC:3235765.3235785}
Plumb, J.N., Kasera, S.K., Stutsman, R.: Hybrid Network Clusters Using Common Gameplay for Massively Multiplayer Online Games. In: Proceedings of the 13th International Conference on the Foundations of Digital Games (2018), pp. 2:1--2:10. ACM. \url{http://doi.acm.org/10.1145/3235765.3235785}

\bibitem{slivar2019qoe}
Slivar, I., Skorin-Kapov, L., Suznjevic, M.: QoE-Aware Resource Allocation for Multiple Cloud Gaming Users Sharing a Bottleneck Link. In: 2019 22nd Conference on Innovation in Clouds, Internet and Networks and Workshops (ICIN) (2019), pp. 118--123.

\bibitem{khmelevsky2017stochastic}
Khmelevsky, Y., Mahasneh, H., Hains, G.: A stochastic gamer's model for on-line games. In: 2017 IEEE 30th Canadian Conference on Electrical and Computer Engineering (CCECE) (2017), pp. 1--4.

\bibitem{mcmanus2019effects}
McManus, J.P., Day, T.G., Mailloux, Z.J.: The Effects of Latency, Bandwidth, and Packet Loss on Cloud-Based Gaming Services (2019). Worcester Polytechnic Institute.

\bibitem{chen2016inter}
Chen, Y., Liu, J., Cui, Y.: Inter-player delay optimization in multiplayer cloud gaming. In: 2016 IEEE 9th International Conference on Cloud Computing (CLOUD) (2016), pp. 702--709.

\bibitem{cai2016survey}
Cai, W.,Shea, R., Huang, C., Chen, K., Liu, J., Leung, V.C.M., Hsu, C.: A survey on cloud gaming: Future of computer games. In: IEEE Access (2016), vol. 4, pp. 7605--7620.

\bibitem{burger2016load}
Burger, V., Pajo, J.F., Sanchez, O.R., Seufert, M., Schwartz, C., Wamser, F., Davoli, F., Tran-Gia, P.: Load dynamics of a multiplayer online battle arena and simulative assessment of edge server placements. In: Proceedings of the 7th International Conference on Multimedia Systems (2016), pp. 17.

\bibitem{Pellegrino2003}
Pellegrino, J.D., Dovrolis, C.: Bandwidth Requirement and State Consistency in Three Multiplayer Game Architectures. In: Proceedings of the 2Nd Workshop on Network and System Support for Games (2003), pp. 52--59. ACM. \url{http://doi.acm.org/10.1145/963900.963905}

\bibitem{Faerber2004}
F{\"a}rber, J.: Traffic modelling for fast action network games. In: Multimedia Tools and Applications (2004), vol. 23 num. 1, pp. 31--46. Springer.

\bibitem{london:2016}
London, I.: Encoding cyclical continuous features - 24-hour time (2016). \url{https://ianlondon.github.io/blog/encoding-cyclical-features-24hour-time/}

\bibitem{chen2016inter}
Chen, Y., Liu, J., Cui, Y.: Inter-player delay optimization in multiplayer cloud gaming. In: 2016 IEEE 9th International Conference on Cloud Computing (CLOUD) (2016), pp. 702--709.

\bibitem{Claypool2006}
Claypool, M., Claypool, K.: Latency and Player Actions in Online Games. In: Commun. ACM (2006), vol. 49 num. 11, pp. 40--45. ACM. \url{http://doi.acm.org/10.1145/1167838.1167860}



%
%
%
%
%


\end{thebibliography}
\end{document}